\documentclass[aps,PRL, a4paper,showkeys,nofootinbib,longbibliography,notitlepage,twocolumn]{revtex4-2}
\usepackage{times}
\usepackage{ulem}
\usepackage{comment}
\usepackage{graphicx}
\usepackage{feynmf}
\usepackage{array}
\usepackage{tabularx}
\usepackage{amsmath}
\usepackage{amstext}
\usepackage{amssymb}
\usepackage{pifont}
\usepackage{xfrac}
\usepackage[colorlinks,citecolor=blue]{hyperref}
\usepackage{graphicx}
\usepackage{amsmath}
\usepackage{amstext}
\usepackage{amssymb}
\usepackage{amsfonts}
\usepackage{longtable,booktabs}
\usepackage{hyperref}
\usepackage{url}
\usepackage{subfigure}
\usepackage{dsfont}
\usepackage{booktabs}
\usepackage{amsbsy}
\usepackage{dcolumn}
\usepackage{amsthm}
\usepackage{bm}
\usepackage{esint}
\usepackage{multirow}
\usepackage{hyperref}
\usepackage{cleveref}
\usepackage{mathrsfs}
\usepackage{amsfonts}
\usepackage{amsbsy}
\usepackage{dcolumn}
\usepackage{bm}
\usepackage{multirow}
\usepackage{color}
\usepackage{extarrows}
\usepackage{datetime}
\usepackage{orcidlink}
\usepackage{mathtools}
\usepackage[super]{nth}
\hypersetup{
	colorlinks=magenta,
	linkcolor=blue,
	filecolor=magenta,
	urlcolor=magenta,
}

\begin{document}
\title{Operator Spreading and Information Propagation: Equivalence and Beyond}

\author{Cheng Shang~\orcidlink{0000-0001-8393-2329}$^{1,2}$}
\email{Contact author: cheng.shang@riken.jp}

\author{Zhi-Guang Lu~\orcidlink{0009-0007-4729-691X}$^{3}$}

\author{Hayato Kinkawa~\orcidlink{0009-0009-5926-5648}$^{2}$}

\author{Tomotaka Kuwahara~\orcidlink{0000-0002-1612-3940}$^{1,4,5}$}

\makeatletter
\renewcommand\frontmatter@affiliationfont{\vspace{1mm} \small}
\makeatother

\affiliation{\textit{$^{1}$Analytical quantum complexity RIKEN Hakubi Research Team, RIKEN Center for Quantum Computing (RQC), Wako, Saitama 351-0198, Japan\\$^{2}$Department of Physics, The University of Tokyo, 5-1-5 Kashiwanoha, Kashiwa, Chiba 277-8574, Japan\\$^{3}$School of Physics, Huazhong University of Science and Technology, Wuhan, 430074, People’s Republic of China \\$^{4}$RIKEN Cluster for Pioneering Research (CPR), Wako, Saitama 351-0198, Japan\\
$^{5}$PRESTO, Japan Science and Technology (JST), Kawaguchi, Saitama 332-0012, Japan}}

\date{\today}

\begin{abstract}
We investigate the quantitative relationship between operator spreading and classical information propagation in quantum systems. Focusing on a bi-partite quantum channel, we derive new upper and lower bounds on the Holevo capacity, a typical information measure, in terms of the trace norm distance between output states, sharpening earlier results by Bravyi \textit{et al}. Our results clarify the extent to which operator growth governs information flow. 
\end{abstract}

\maketitle

\textit{Introduction}. Operator spreading describes how, initially, local operators evolve to become non-local under unitary (or dissipative) dynamics in quantum many-body systems. This phenomenon has become a central concept in various areas of physics, including studies of quantum chaos, thermalization, and even black hole information paradoxes~\cite{PhysRevE.50.888,Hayden_2007,PhysRevX.8.021014,Maldacena_2020,PhysRevB.108.224306,Haehl_2023}.~A standard way to quantitatively characterize operator spreading is through the Lieb-Robinson (LR) bound~\cite{_Lieb_Robinson_1972,Nachtergaele_2006}, which sets a fundamental speed limit for information propagation in non-relativistic systems. Closely related to this is the out-of-time-order correlator (OTOC)~\cite{Maldacena_2016}, which serves as a widely used diagnostic for operator spreading and quantum information scrambling~\cite{_Swingle_2018,PhysRevLett.126.030604,PhysRevB.107.205127,PRXQuantum.5.010201}.

From an information-theoretic perspective~\cite{PhysRevB.96.020406,PhysRevLett.134.020201}, operator spreading can be linked to the concept of information propagation. If an operator fails to spread, then any information encoded locally (e.g., via a local CPTP map) cannot be detected in distant regions.
Thus, it is natural to expect that the absence of operator spreading implies the absence of information transfer. This intuition is supported by prior work~\cite{PhysRevLett.97.050401}, which provided a natural and simple upper bound relating operator spreading to the Holevo capacity, the maximum amount of classical information that can be transmitted through a quantum channel~\cite{_Holevo_1998}.
While this result captures an essential aspect of the connection, it only constrains information propagation from above and does not offer a complete characterization. A more refined understanding, particularly one that includes both upper and lower bounds, remains an open question.


In this work, we aim to probe this relation more deeply by asking a fundamental question: \textit{Is there a more direct correspondence between information propagation and operator spreading?} To this end, we consider a simple setting where an operator initially supported on a subsystem \( A \) spreads to another subsystem \( B \) under time evolution. The central question is whether such spreading is both a necessary and sufficient condition for transmitting information from \( A \) to \( B \).
Indeed, seminal work by Beckman, Gottesman, Nielsen, and Preskill~\cite{PhysRevA.64.052309} rigorously established the \textit{equivalence} between the absence of information propagation and the absence of operator spreading, provided that the underlying dynamics is governed by a completely positive and trace-preserving (CPTP) map. Specifically, they showed that no information can be transmitted between two parties if and only if the CPTP map factorizes as a tensor product of local operations.
However, while this result provides a qualitative connection, a general \textit{quantitative} relationship between operator spreading—characterized by the commutator norm \( \| [O_A(t), O_B] \| \)—and the amount of transmittable information—captured by the Holevo capacity—has not been thoroughly explored. In particular, our study seeks to determine how tightly these two quantities are connected and whether one can derive meaningful, general bounds that link them across different dynamical regimes.

This work aims to take a more precise step toward understanding the quantitative relationship between operator spreading and the Holevo capacity.
In particular, we revisit a foundational result by Bravyi \textit{et al}.~\cite{PhysRevLett.97.050401}, which established an upper bound on the Holevo capacity in terms of operator spreading. 
Focusing on the setting of the quantum channel between two subsystems \( A \) and \( B \), we refine this result by deriving a tighter and more general inequality that provides an upper bound on the Holevo capacity as a function of the commutator norm \( \| [O_A(t), O_B] \| \).
These findings not only sharpen the mathematical connection between operator spreading and information propagation but also clarify the extent to which the two notions are quantitatively linked in general quantum systems.

This Letter is organized as follows. We first introduce the setup, then derive new bounds, and finally conclude with implications and future directions.

\textit{Setup}. To investigate how information propagates during operator spreading, we rigorously quantify the relationship between operator spreading and the Holevo capacity. 

To clarify the relation between operator spreading and information propagation, we consider a bipartite quantum system composed of two subsystems, \( A \) and \( B \). We set the global initial state of the system to be \( \rho_0^{AB} \). A local encoding $\mathcal{E}_A$ is first applied to it, and the encoded quantum state subsequently undergoes time evolution governed by a bipartite CPTP map $\tau_{AB,t}$ (see below).

To model the process of local encoding, we introduce a classical ensemble \( \mathcal{E}_A = \{ p_i,{U_{A,i}}\} \), where each ${U_{A,i}}$ is a unitary operation acting only on subsystem \( A \). Given this ensemble, the encoded states are defined as 
\begin{align}
\rho_i^{AB} &= (U_{A,i} \otimes \mathbb{I}_B) \rho_0^{AB} (U_{A,i}^\dagger \otimes \mathbb{I}_B),
\end{align}
each occurring with probability \( p_i \). After time evolution, the reduced state of subsystem \( B \) becomes
\begin{equation}
\rho _i^B(t) = {{\mathop{\rm Tr}\nolimits} _A}\left[ {{\tau_{AB,t}}\rho _i^{AB}} \right].
\end{equation}
Here, $\tau_{AB,t}$ is a CPTP map acting on the joint system $AB$, and hence it can be represented in the Kraus form as
\begin{align}
{\tau _{AB,t}}\rho _i^{AB} = \sum\limits_\alpha  {{K_{\alpha ,t}}} \rho _i^{AB}K_{\alpha ,t}^\dag ,
\end{align}
where the Kraus operators $\left\{ {{K_{\alpha ,t}}} \right\}$ satisfy the completeness relation: $\sum\nolimits_\alpha  {K_{\alpha ,t}^\dag {K_{\alpha ,t}}}  = {\mathbb{I}_{{AB}}}$.

\smallskip
The Holevo capacity, which quantifies the maximum amount of classical information transmittable from \( A \) to \( B \), is defined as~\cite{_Holevo_1998}
\begin{align}
C_\chi(t) = S^{B}\left( \sum_i p_i \rho_i^B(t) \right) - \sum_i p_i S^{B}\left( \rho_i^B(t) \right), \label{HolevoCapacitySimple}
\end{align}
where \( S^B(\cdot) \) denotes the von Neumann entropy.

\smallskip
To further bound the Holevo capacity, we consider the trace-norm difference between the reduced states before and after encoding. For any bounded local operator \( O_B \) with \( \| O_B \| \le 1 \), the following inequality holds~\cite{PhysRevLett.97.050401}:
\begin{align}
\left\| \rho_i^B(t) - \rho_0^B(t) \right\|_1 
&\le \sup_{\| O_B \| = 1} \left\| \left[ {U_{A,i}}, O_B(t) \right] \right\| \ \ \forall i, \label{eq:trace-norm-vs-commutator}
\end{align}where ${\left\|  \cdot  \right\|_1}$ and $\left\|  \cdot  \right\|$ represent the trace norm and the operator norm, respectively. Here, $\rho_0^B(t)$ corresponds to the state with no encoding onto $A$, and $O_B(t)$ denotes the time-evolved operator in the Heisenberg picture, which can be expressed as
\begin{align}
{O_B}(t) = \tau _{AB,t}^\dag {O_B} = \sum\limits_\alpha  {K_{\alpha ,t}^\dag } {O_B}{K_{\alpha ,t}}.
\end{align} Inequalities~(\ref{eq:trace-norm-vs-commutator}) connects the operator spreading, as measured by the growth of commutators, to the trace-norm distinguishability of reduced states on subsystem \( B \). Hence, operator spreading directly constrains the amount of information about local encodings that can be accessed from \( B \) at time \( t \).
\smallskip

Without loss of generality, as discussed above, we assume that \( A \) encodes information by applying a set of unitary operations \( \{ U_{A,i} \} \), where \( i \) varies depending on the message that \( A \) intends to send to \( B \). To set the stage and generalize, we recall a key structural result that holds even beyond the unitary setting. In general, any local operation $\{ \tau_{A,i} \}$ on $A$ can be described by a completely positive and trace-preserving (CPTP) map, which can be implemented via a unitary evolution on an enlarged system, where an ancillary system is incorporated into $A$~\cite{PhysRevA.76.052319}. Reference~\cite{PhysRevA.64.052309} proved that, under CPTP maps, information propagation is possible if and only if the map fails to factorize into a tensor product of local operations. This result implies that the absence of operator spreading corresponds exactly to the absence of accessible information transmission:
\smallskip

\noindent
\textbf{Claim 1 (Known result).}~
\textit{The Holevo capacity equals zero for any encoding \( \mathcal{E}_A = \{ p_i, U_{A,i} \} \) if and only if the quantum channel is a product map:}
\begin{align}
 C_\chi(t) = 0 \quad \Longleftrightarrow \quad {\tau _{AB,t}} = {\tau _{A,t}} \otimes {\tau _{B,t}}. \label{Claim 1}
\end{align}\textit{In other words, for any non-product quantum channel, there always exists an encoding that enables information transmission from \( A \) to \( B \).}
\smallskip

This equivalence establishes a minimal baseline: if no information reaches subsystem \( B \), then all output states $\rho _i^B\left( t \right)$ must coincide. Conversely, any nonzero Holevo capacity implies distinguishability and hence operator spreading.
\smallskip

We now go beyond this qualitative link to derive quantitative bounds on the Holevo capacity regarding operator-theoretic measures.

\smallskip
\textit{Quantifying the relationship}. After Claim~1 established the necessary and sufficient condition for identifying accessible information propagation, our main result quantifies the relationship between the Holevo capacity and operator spreading. As a central result, we derive new upper and lower bounds on the Holevo capacity and demonstrate that the upper bound is directly related to operator spreading, thereby providing a quantitative bridge between dynamics and information flow in quantum systems.

\begin{widetext}
\noindent \textbf{Theorem 1.}~\textit{By employing the quantum skew divergence and defining the complementary state associated with the reduced density matrix $\rho_i^B(t)$ as $\tilde{\rho}_i^B(t)$, the time-dependent Holevo capacity satisfies the following inequality:}
\begin{align}
   \sum_i \frac{1}{2}p_i (1 - p_i)^2 \left\| \rho_i^B(t) - \tilde{\rho}_i^B(t) \right\|_1^2
   \le C_\chi(t) \le - \sum_i \frac{1}{2} p_i \log(p_i) \left\| \rho_i^B(t) - \tilde{\rho}_i^B(t) \right\|_1,
   \label{time-dependent-Holevo}
\end{align}
\textit{where the complementary state is defined as}
\begin{align}
\tilde{\rho}_i^B(t) := \frac{\bar{\rho}^B(t) - p_i \rho_i^B(t)}{1 - p_i},
\end{align}
\textit{with $\bar{\rho}^B(t) := \sum_i p_i \rho_i^B(t)$ being the ensemble average.}
\end{widetext}

To support the proof, we first introduce the quantum skew divergence~\cite{Audenaert_2014}, a measure defined for density operators $\eta$ and $\mu$ by
\begin{align}
\label{Skew_divergence}
S_\lambda(\eta \| \mu) := \frac{1}{- \log \lambda} S\left( \eta \middle\| \lambda \eta + (1 - \lambda)\mu \right),
\end{align}
where $0 < \lambda < 1$, and $S(\cdot \| \cdot)$ denotes the quantum relative entropy. This divergence satisfies~\cite{Temme_2010,Audenaert_2014}
\begin{align}
\frac{2 (1 - \lambda)^2}{- \log \lambda} \, T(\eta, \mu)^2 \le S_\lambda(\eta \| \mu) \le T(\eta, \mu),
\end{align}
where $T(\eta, \mu) := \frac{1}{2} \| \eta - \mu \|_1$ is the trace distance.
\medskip

\textit{Proof of Theorem 1.}  
We first rewrite the Holevo capacity as
\begin{align}
C_\chi(t) = \sum_i p_i S^{B}\left( \rho_i^B(t) \middle\| \bar{\rho}^B(t) \right).
\end{align}
Using the identity $\bar{\rho}^B(t) = p_i \rho_i^B(t) + (1 - p_i) \tilde{\rho}_i^B(t)$, the relative entropy can be re-expressed in terms of the skew divergence:
\begin{align}
S^{B}\left( \rho_i^B(t) \middle\| \bar{\rho}^B(t) \right) = - \log p_i \cdot S^{B}_{p_i}\left( \rho_i^B(t) \middle\| \tilde{\rho}_i^B(t) \right),
\end{align}
leading to
\begin{align}
C_\chi(t) = - \sum_i p_i \log p_i \cdot S^{B}_{p_i}\left( \rho_i^B(t) \middle\| \tilde{\rho}_i^B(t) \right),
\end{align}
where we apply $\lambda=p_i$ in the definition~\eqref{Skew_divergence}.
Then, applying the skew divergence bounds yields the desired inequality in Eq.~\eqref{time-dependent-Holevo}. $\square$
\medskip

\textit{Discussion.} (i) The bounds in Eq.~\eqref{time-dependent-Holevo} are physically meaningful and always lie within $[0, H(\mathbf{P})]$, where $H(\mathbf{P}) = -\sum_i p_i \log p_i$ is the Shannon entropy of the input ensemble. 
(ii) Compared to the standard upper bound $C_\chi \le H(\mathbf{P})$~\cite{_Nielsen_Chuang_2010}, our result is strictly tighter whenever operator spreading is limited. (iii) The trace distance can be bounded in a state-independent manner. Let $T_{\max} := \max_{i,j} T(\rho_i^B, \rho_j^B)$. Then,
\begin{align}
C_\chi(t) \le H(\mathbf{P}) \cdot T_{\max}.
\end{align}

\noindent (iv) In a quantum many-body setup, if the time evolution satisfies the Lieb-Robinson bound, we can derive
\begin{align}
C_\chi(t) \le \frac{1}{2} H(\mathbf{P})\varepsilon _{\rm{LR}},
\end{align}
where ${\varepsilon _{\rm{LR}}}$ is determined by the LR bound in the system. This bound captures the light-cone-like structure of information spreading. (v) Unlike prior results~\cite{PhysRevLett.97.050401}, which involve system dimension, our bound is dimension-independent and thus more broadly applicable in many-body contexts.
\smallskip

\textit{Summary and outlook}. In this work, we have refined the relationship between operator spreading and information propagation through the framework of the Holevo capacity. By developing a series of general inequalities and structural characterizations, we have clarified how and to what extent operator spreading constrains the ability of quantum dynamics to transfer classical information.

Our main result, Theorem~1, establishes upper and lower bounds on the Holevo capacity in terms of trace-norm differences between output states. Importantly, the derivation does not rely on unitary dynamics and thus holds for general quantum channels described by completely positive trace-preserving (CPTP) maps. This demonstrates that operator spreading, when appropriately defined (e.g., through commutator growth or trace distances in the Heisenberg picture), continues to provide a quantitative handle on classical information propagation for general quantum channels.

Another important scenario arises when the system is extended to a tripartite configuration $ABC$, where information is transmitted from $A$ to $B$ through the intermediate system $C$. In such settings, the operator initially localized in $A$ may spread across the entire system $ABC$, making the relationship between operator spreading and the Holevo capacity significantly more opaque. In particular, the conditions under which the Holevo capacity vanishes become much harder to characterize, and no clear necessary and sufficient criteria are currently known.
This type of mediated communication is not merely of theoretical interest, as it frequently occurs in realistic models of quantum many-body systems, where information often propagates indirectly through intermediate subsystems. Understanding the interplay between multipartite operator spreading and information transmission in such setups remains a largely unexplored but highly promising direction for future research.

\smallskip
\textit{Acknowledgments}. C. S. thanks Naomichi Hatano, Tan Van Vu, Zongping Gong, Takano Taira, Miku Ishizaki, Jaeha Lee, and Donghoon Kim for their valuable discussions. C. S. also extends special thanks to the committee members of the Ph.D. defense, Yuto Ashida, Masahito Ueda, Mio Murao, Kiyotaka Aikawa, and Kuniaki Konishi, for their insightful feedback and constructive comments. C. S. acknowledges financial support from the China Scholarship Council, the Japanese Government (Monbukagakusho-MEXT) Scholarship (Grant No.~211501), and the RIKEN Junior Research Associate Program. C. S. and T. K. acknowledge the Hakubi Projects of RIKEN. T. K. was also supported by JST PRESTO (Grant No.~JPMJPR2116), ERATO (Grant No.~JPMJER2302), and JSPS Grants-in-Aid for Scientific Research (Grant No.~JP23H01099 and JP24H00071), Japan. H. K. was supported by JST SPRING (Grant No.~JPMJSP2108).

\bibliography{reference}

\end{document}